\documentclass[12pt]{article}
 \usepackage{epsfig}
 \def\be{\begin{equation}}
 \def\ee{\end{equation}}
 \def\bea{\begin{eqnarray}}
 \def\eea{\end{eqnarray}}
 \usepackage{graphicx}

 \catcode`\@=11
 \def\lsim{\mathrel{\mathpalette\@versim<}}
 \def\gsim{\mathrel{\mathpalette\@versim>}}
 \def\@versim#1#2{\vcenter{\offinterlineskip
 \ialign{$\m@th#1\hfil##\hfil$\crcr#2\crcr\sim\crcr } }}
 \catcode`\@=12

 \catcode`@=12
\begin{document}
 \thispagestyle{empty}
 \begin{flushright}
 UCRHEP-T576\\
 April 2017\
 \end{flushright}
 \vspace{0.6in}
 \begin{center}
 {\LARGE \bf Inception of Self-Interacting Dark Matter\\ with 
Dark Charge Conjugation Symmetry\\}
 \vspace{1.2in}
 {\bf Ernest Ma\\}
 \vspace{0.2in}
 {\sl Physics and Astronomy Department,\\ 
 University of California, Riverside, California 92521, USA\\}
 \end{center}
 \vspace{1.2in}

\begin{abstract}\
A new understanding of the stability of self-interacting dark matter is 
pointed out, based on the 
simplest spontaneously broken Abelian $U(1)$ gauge model with one complex 
scalar and one Dirac fermion.  The key is the imposition of dark charge 
conjugation symmetry.  It allows the possible existence of two stable 
particles: the Dirac fermion and the vector gauge boson which acts 
as a light mediator for the former's self-interaction.  Since this light 
mediator does not decay, it avoids the strong cosmological constraints 
recently obtained for all such models where the light mediator decays 
into standard-model particles. 
\end{abstract}

\newpage
\baselineskip 24pt
\noindent \underline{\it Introduction}~:~
The Lagrangian of the simplest spontaneously broken Abelian $U(1)$ gauge model 
was written down by Peter Higgs over 50 years ago~\cite{h64}.  Its particle 
content consists of a vector gauge boson (call it $Z_D$) and a complex scalar 
(call it $\sigma$).  By itself it has automatic charge conjugation 
invariance, i.e. $Z_D \to -Z_D$, $\sigma \to \sigma^*$, resulting in 
$g_D \to -g_D$. After spontaneous symmetry breaking, the above still holds, 
i.e. $Z_D \to -Z_D$, $\sigma_R \to \sigma_R$, and $\sigma_I \to -\sigma_I$ 
which becomes the longitudinal component of the now massive $Z_D$.  This fact 
has been used~\cite{fr12,bkps13,dft16,dm17} to suggest that $Z_D$ may be 
dark matter. 

The existence of two $U(1)$ gauge factors allows for the gauge-invariant 
kinetic mixing~\cite{h86} of the two associated gauge bosons, so $Z_D$ may 
mix with the $U(1)_Y$ gauge boson of the standard model (SM), of which the 
photon is a component.  This has led to many theoretical studies of a possible 
light dark photon, and the experiments which may be relevant in finding 
it~\cite{dark16}.  However, this kinetic mixing term breaks the dark charge 
conjugation symmetry, so the former may be absolutely forbidden if  
the latter is chosen to be exact.

In the Higgs model, $Z_D$ is the sole dark matter.  Suppose a Dirac 
fermion (call it $N$) is added, transforming also under $U(1)_D$, 
then the Lagrangian is also invariant under dark charge conjugation, as 
well as the global $U(1)$ transformation operating on $N$, i.e. dark fermion 
number.  Hence $N$ is a dark-matter candidate.  What about $Z_D$?  If 
$m_{Z_D} > 2 m_N$, then $Z_D$ will decay into $N \bar{N}$ through the vector 
current $\bar{N} \gamma_\mu N$ which has charge conjugation ${\cal C}=-1$, 
but if  $m_{Z_D} < 2 m_N$, then $Z_D$ will be stable.  Further, if 
$Z_D$ is much lighter than $N$, then it may act as a stable light 
mediator for $N$ self-interactions. 
Note that if $Z_D$ is unstable and decays to SM particles, then very 
strong constraints exist~\cite{bksw17} which basically rule out this 
scenario for explaining~\cite{fkty09} the core-cusp anomaly observed 
in dwarf galaxies~\cite{dgs09}.  As for the dark Higgs boson 
$h_D = \sqrt{2} Re(\sigma)$, it may also be light, but it has an 
unavoidable mixing with the SM Higgs boson $h$, so it will not be 
stable.  In the following, $m_{h_D} < m_{Z_D}$ will be assumed.

With $m_N \sim 100$ GeV and $m_{Z_D} \sim 10$ MeV, the $N \bar{N}$ 
annihilation to $Z_D Z_D$ is assumed to have the right cross section 
for $N$ to be the main component of dark matter.  The subsequent $Z_D Z_D$ 
annihilation to $h_D h_D$ is assumed to have a large enough cross section, 
so that the relic abundance of $Z_D$ is small compared to that of $N$.  
In direct-search experiments, $N$ does not interact with quarks, so 
there will be no signal.  As for the small $Z_D$ component, it interacts 
through $h_D-h$ mixing, but since $Z_D$ is very light, current experiments 
are not sensitive to its presence.  On the other hand, the $h_D-h$ mixing 
has to be large enough for it to decay away before big bang nucleosynthesis 
(BBN).  Even so, $h_D$ may be produced at late times through $Z_D Z_D$ 
annihilation, and affects the cosmic microwave background (CMB) through its 
decay, as pointed out in Ref.~\cite{bksw17}.  However, there is no 
Sommerfeld enhancement~\cite{s31} of this cross section, unlike the case 
of $N \bar{N}$ 
annihilation through a light mediator which decays.  Hence the proposed model 
is a natural resolution of this conundrum, as detailed below.

\noindent \underline{\it Dark $U(1)_D$ model}~:~
This model assumes $U(1)_D$ gauge symmetry, implying thus a vector gauge 
boson $Z_D$.  It is spontaneously broken by a complex scalar $\sigma$ 
with charge $g_D$.  A Dirac fermion $N$ also exists with charge $g_N$. 
The complete Lagrangian before symmetry breaking is
\begin{eqnarray}
{\cal L} &=& -{1 \over 4} (\partial^\mu Z_D^\nu - \partial^\nu Z_D^\mu) 
(\partial_\mu Z_{D\nu} - \partial_\nu Z_{D\mu}) + (\partial^\mu \sigma - 
ig_D Z_D^\mu \sigma)(\partial_\mu \sigma^* + i g_D Z_{D\mu} \sigma^*) \nonumber \\ 
&+& \mu_D^2 \sigma^* \sigma - {1 \over 2} \lambda_D (\sigma^* \sigma)^2 
+ i \bar{N} \gamma_\mu (\partial^\mu - ig_N Z_D^\mu) N - m_N \bar{N} N.
\end{eqnarray}
In the above, if we replace $g_D$ by $-g_D$, $\sigma$ by $\sigma^*$, 
$g_N$ by $-g_N$, and $N$ by its dark charge conjugate, we have exactly 
the same physical theory.  The spontaneous breaking of $U(1)_D$ with 
$\langle \sigma \rangle = v_D/\sqrt{2}$ changes the Lagrangian to
\begin{eqnarray}
{\cal L} &=& -{1 \over 4} (\partial^\mu Z_D^\nu - \partial^\nu Z_D^\mu) 
(\partial_\mu Z_{D\nu} - \partial_\nu Z_{D\mu}) + {1 \over 2} m^2_{Z_D} Z_D^\mu 
Z_{D\mu} + {1 \over 2}(\partial^\mu h_D)(\partial_\mu h_D) - {1 \over 2} 
m^2_{h_D} h_D^2 \nonumber \\ &+& {m^2_{h_D} \over 2 v_D} h_D^3 + 
{m^2_{h_D} \over 8 v_D^2} h_D^4 + g_D^2 v_D h_D (Z^\mu_D Z_{D\mu}) 
+ {1 \over 2} g_D^2 h_D^2 (Z^\mu_D Z_{D\mu}) \nonumber \\ &+& 
i \bar{N} \gamma_\mu \partial^\mu N - m_N \bar{N} N + g_N Z^\mu_D \bar{N} 
\gamma_\mu N,
\end{eqnarray}
where $v_D^2 = 2 \mu_D^2/\lambda_D$, $m_{Z_D} = g_D v_D$, and $m^2_{h_D} = 
\lambda_D v_D^2$.  The crucial interaction terms are $g_N Z^\mu_D \bar{N} 
\gamma_\mu N$, $g_D^2 v_D h_D (Z^\mu_D Z_{D\mu})$, and $(1/2) g_D^2 h_D^2 
(Z^\mu_D Z_{D\mu})$.  We assume in the following $m_N \sim 100$ GeV, with 
$Z_D,h_D \sim 10$ MeV, with $m_{h_D} < m_{Z_D}$.  Note that $g_N$ is 
independent of $g_D$.

\noindent \underline{\it Three new particles}~:~
There are only three new particles beyond those of the standard model. 
Each serves a purpose and is an essential ingredient of this two-component 
dark-matter model.  The dark fermion $N$ is a Dirac particle with a conserved 
dark fermion number.  It is the dominant component of the observed dark 
matter of the Universe.  It has a dark gauge interaction mediated by $Z_D$ 
which is light, thus realizing the requirement of a sufficiently large 
interaction to affect the core-cusp discrepancy of dwarf galaxies. 
The imposition of dark charge conjugation symmetry means that $Z_D$ has 
${\cal C} = -1$.  It couples to the vector current $\bar{N} \gamma_\mu N$ 
which also has ${\cal C} = -1$, so it may decay into $N \bar{N}$, but 
if it is lighter than $2 m_N$ as assumed, then it is itself stable. 
As such, it may be overproduced in the early Universe.  However, it is 
also assumed that the dark Higgs boson $h_D$, which breaks the $U(1)_D$ 
gauge symmetry and provides $Z_D$ with a mass through its vacuum expectation 
value $v_D$, is lighter than $Z_D$.  Hence the $Z_D Z_D \to h_D h_D$ 
annihilation should be strong enough to make it a very small fraction of the 
observed dark matter of the Universe.  As for $h_D$, which has ${\cal C} 
= +1$, it must be unstable through its allowed mixing with the SM Higgs 
boson $h$, and decays away early without affecting the standard BBN.  

Consider the extended scalar potential involving both $\sigma$ and the 
SM Higgs doublet $\Phi = (\phi^+,\phi^0)$:
\begin{eqnarray}
V &=& -\mu_D^2 \sigma^* \sigma + {1 \over 2} \lambda_D (\sigma^* \sigma)^2 
- \mu_h^2 \Phi^\dagger \Phi + {1 \over 2} \lambda_h (\Phi^\dagger \Phi)^2 
+ \lambda_{hD} (\sigma^* \sigma)(\Phi^\dagger \Phi).
\end{eqnarray}
Using $\phi^0 = (v_h+h)/\sqrt{2}$, the $2 \times 2$ mass-squared 
matrix spanning $(h_D,h)$ is given by
\begin{equation}
{\cal M}^2_{h_D,h} = \pmatrix{\lambda_D v_D^2 & \lambda_{hD} v_D v_H \cr 
\lambda_{hD} v_D v_H & \lambda_h v_h^2}.
\end{equation}
Assuming $m_{h_D} << m_{h} = 125$ GeV, the $h_D-h$ mixing is then 
$\theta_{hD} = \lambda_{hD} v_D v_h/m^2_h$.  For a light $h_D$ of order 
10 MeV, its dominant decay is to $e^- e^+$ with the decay rate
\begin{equation}
\Gamma (h_D \to e^-e^+) = {m_{h_D} m_e^2 \over 8 \pi v_h^2} \theta^2_{hD},
\end{equation}
where $v_h = 246$ GeV. 
Assuming that $\Gamma^{-1} < 1~s$, the constraint
\begin{equation}
\left( {m_{h_D} \over 10~{\rm MeV}} \right) \theta^2_{hD} > 
3.83 \times 10^{-10}
\end{equation}
is obtained.  The SM Higgs boson $h$ also decays into $h_D h_D$ with 
coupling $\lambda_{hD} v_h$.  Its decay rate is
\begin{equation}
\Gamma (h \to h_D h_D) = {\lambda_{hD}^2 v_h^2 \over 16 \pi m_h} = 
\lambda_{hD}^2 (9.63~{\rm GeV}).
\end{equation}
Assuming that this is no more than 10$\%$ of the Higgs boson width in the SM 
(4.12 MeV), this gives a bound of 
\begin{equation}
\lambda_{hD} < 0.0066.
\end{equation}
Comparing Eqs.~(6) and (7), the constraint
\begin{equation}
\left( {v_D \over {\rm GeV}} \right) > 0.19 \sqrt{{\rm 10~MeV} \over m_{h_D}}
\end{equation}
is obtained.

\noindent \underline{\it $Z_d Z_d$ annihilation}~:~
Consider first the process $Z_d Z_d \to h_D h_D$ at rest.  There are 
four diagrams summing up to the amplitude
\begin{equation}
{\cal A} = \left[ {2 g_D^2 (2 + r) \over 2-r} - {6 g_D^2 r \over 4-r} 
\right] (\vec{\epsilon}_1 \cdot \vec{\epsilon}_2) + {8 g_D^2 \over m^2_{Z_D} 
(2-r)} (\vec{\epsilon}_1 \cdot \vec{k})(\vec{\epsilon}_2 \cdot \vec{k}),
\end{equation}
where $r = m^2_{h_D}/m^2_{Z_D}$ and the center-of-mass variables $\vec{k}$ 
(momentum of $h_D$) and $\vec{\epsilon}_{1,2}$ (polarizations of $Z_D$) 
have been used.  The resulting cross section $\times$ relative velocity 
is given by
\begin{equation}
\sigma(Z_D Z_D \to h_D h_D) \times v_{rel} = {g_D^4 \sqrt{1 - r} \over 
64 \pi m^2_{Z_D}} \left[ {4 [r^2 + 4(2-r)^2] \over (4-r)^2} - {24 r (2+r) 
\over 9(2-r)(4-r)} + {8 (2+r)^2 \over 9(2-r)^2} \right].
\end{equation}
Let $m_{Z_D} = 10$ MeV and $m_{h_D} = 8$ MeV, then $r=0.64$.  The coupling 
$g_D$ is adjustable.  Let $g_D = 0.005$ for example, then
\begin{equation}
\sigma(Z_D Z_D \to h_D h_D) \times v_{rel} = 1.1 \times 10^{-24}~{cm}^3/s, 
\end{equation}
which is 37 times the canonical $\sigma_0 \times v_{rel} = 3 \times 
10^{-26}~cm^3/s$ for obtaining the correct dark-matter relic abundance 
of the Universe.   This means that $Z_D$ will be underproduced and forms 
only a small component of the observed dark matter, which will be mainly 
$N$ as discussed in the next section.  Note also that $g_D = 0.005$ and 
$m_{Z_D} = 10$ MeV imply that $v_D = 2$ GeV, which is perfectly consistent 
with Eq.~(9).

\noindent \underline{\it $N \bar{N}$ annihilation}~:~
The annihilation $N \bar{N} \to Z_D Z_D$ is analogous to $e^- e^+ \to \gamma 
\gamma$.  The cross section at rest $\times$ relative velocity is given by
\begin{equation}
\sigma(N \bar{N} \to Z_D Z_D) \times v_{rel} = {g_N^4 \over 16 \pi m_N^2}.
\end{equation}
For $m_N = 100$ GeV, this would be equal to $2 \sigma_0 \times v_{rel} = 
6 \times 10^{-26}~cm^3/s$ if $g_N = 0.225$.  For the light mediator 
with $m_{Z_D} = 10$ MeV, Sommerfeld enhancement is expected.  
However, at the time of thermal freezeout, this effect is only 
${\cal O}(1)$~\cite{abp12,tyz13}.  The large enhancement will come at late 
times (because of the decreasing relative velocity of $N \bar{N}$ 
annihilation) and may be as large as a factor of $10^4$.  Whereas the 
fraction of $N \bar{N}$ which would annihilate is still negligible 
compared to the entire population, the production of an unstable mediator 
would allow its decay products (photons and electrons) to affect the 
CMB, thus ruling out (for $s$-wave annihilation) all models where the 
self-interactions are large enough to address the small-scale problems of 
structure formation, as pointed out recently~\cite{bksw17}.

Here the light mediator $Z_D$ is stable, so it does not affect the CMB.  
As for $h_D$, it may also be produced at late times from $Z_D Z_D$ 
annihilation, but this cross section has no Sommerfeld enhancement, 
so even though $h_D$ decays to $e^- e^+$, its effect is small.

\noindent \underline{\it Thermal history}~:~
The dark fermion $N$ is kept in thermal equilibrium with its light mediator 
$Z_D$ which couples to the dark Higgs boson $h_D$.  The bridge connecting 
the dark sector with the SM is the quartic scalar interaction term 
$\lambda_{hD} (\sigma^* \sigma)(\Phi^\dagger \Phi)$ of Eq.~(3).  Hence 
$h_D$ is in thermal equilibrium with the SM Higgs boson $h$, and through the 
latter, all the SM particles.  As the Universe cools below $m_N$, $N$ freezes 
out with a relic abundance which accounts for most of the observed dark 
matter of the Universe. 
In structure formation, $N$ has a large enough elastic cross section due to 
the exchange of its light mediator $Z_D$ to explain the flatter density 
profiles of dwarf galaxies near their centers~\cite{fkty09}.

The light vector boson $Z_D$ is stable and interacts with $h_D$ to remain in 
thermal equilibrium until the Universe cools below $m_{Z_D}$.  It then 
freezes out with a much smaller relic abundance than that of $N$. The 
dark Higgs boson $h_D$ decays away quickly at early times through its mixing 
with the SM Higgs boson $h$.  All these happen before the onset of BBN 
so that the standard predictions of all relevant 
cosmological parameters are unchanged.  At late times, $Z_D$ re-emerges 
from $N \bar{N}$ annihilation, but it is stable and will not disturb the 
CMB.  The dark Higgs boson $h_D$ also 
re-emerges from $Z_D Z_D$ annihilation, but this cross section is not 
enhanced by the Sommerfeld effect, so even though $h_D$ decays to 
$e^-e^+$, its effect on the CMB is harmless.

\noindent \underline{\it Phenomenological consequences}~:~
The model presented has a dark gauge $U(1)_D$ symmetry, with exact dark 
charge conjugation invariance.  It has two stable particles, the dark 
fermion $N$ with $m_N \sim 100$ GeV and a light vector mediator $Z_D$ 
with $m_{Z_D} \sim 10$ MeV.  As such, it explains the observed relic 
abundance of dark matter, as well as the cusp-core anomaly of dwarf 
galaxies.  It avoids the strong constraints of decaying 
particles on the CMB~\cite{bksw17}.  The $U(1)_D$ symmetry is broken 
with $v_D \sim 2$ GeV 
as constrained by Eq.~(9).  The associated dark Higgs boson $h_D$ is 
lighter than $Z_D$ and mixes with the SM Higgs boson $h$.

In direct-search experiments, $N$ is essentially invisible because it has 
only $Z_D$ interactions which do not affect SM particles at tree level.  
As for $Z_D$, its relic abundance is suppressed and its mass is only 
about 10 MeV, so even though it interacts with SM particles through 
$h_D-h$ mixing, it is insensitive to present underground experiments. 
This would not be the case if $m_{Z_D} \sim 100$ GeV.  In fact, it has been 
shown~\cite{kty14} that a light mediator would then be ruled out because the 
direct-detection bound excludes its decay before the onset of BBN. 
In indirect-search experiments, the $N \bar{N}$ annihilation is 
Sommerfeld-enhanced, but it only produces $Z_D$ at tree level which 
cannot be detected.  In one loop, SM particles may be produced, but 
the cross section is very small.  Hence neither types of the 
conventional search for dark matter would have much promise in 
detecting such dark matter.

Since the light vector boson $Z_D$ has no kinetic mixing with the photon 
because of the dark gauge conjugation symmetry, there is also no effect 
on experiments searching for it through this portal.  

A possible way to discover $h_D$ is from $h \to h_D h_D$ decay at an 
accelerator, and the subsequent decay $h_D \to e^-e^+$.  The problem is 
that $h_D$ has a lifetime of about 1 $s$, so the decay products are far 
downstream and not easily observed.

\noindent \underline{\it Remarks}~:~
The idea of self-interacting dark matter is faced with a 
conundrum~\cite{bksw17}.   If the interaction is strong enough to 
address the small-scale problems of structure formation, the production 
of the light mediator at late times would disrupt the cosmic microwave 
background because of the inherent Sommerfeld enhancement for $s$-wave 
annihilation and the apparently inescapable fact that the mediator must 
decay into electrons or photons.  Its resolution in terms of a simple 
complete renormalizable model is the subject matter of this paper. 
Unfortunately, this model predicts null or negligible effects 
in all present attempts to discover the nature of dark matter. 
On the other hand, it may be the answer to the question of why 
dark matter has not been seen so far.

\newpage
\noindent \underline{\it Acknowledgement}~:~
This work was supported in part by the U.~S.~Department of Energy Grant 
No. DE-SC0008541.

\baselineskip 18pt

\bibliographystyle{unsrt}

\end{document}